\documentclass[reprint, superscriptaddress, showpacs, amsmath, amssymb, aps, prl]{revtex4-1}

\usepackage{graphicx}
\usepackage{dcolumn}
\usepackage{bm}
\usepackage{hyperref}
\usepackage[mathlines]{lineno}
\usepackage[caption=false]{subfig}
\pdfoutput=1

\begin{document}

\preprint{Hose mitigation}

\title{Mitigation of the hose instability in plasma-wakefield accelerators}
\author{T.J.~Mehrling}\email{timon.mehrling@desy.de}
\affiliation{Institut f\"ur Experimentalphysik, Universit\"at Hamburg, 22761 Hamburg, Germany}
\affiliation{GoLP/Instituto de Plasmas e Fus\~ao Nuclear, Instituto Superior T\'ecnico, Universidade de Lisboa, 1049-001 Lisboa, Portugal}
\author{R.A.~Fonseca}
\affiliation{GoLP/Instituto de Plasmas e Fus\~ao Nuclear, Instituto Superior T\'ecnico, Universidade de Lisboa, 1049-001 Lisboa, Portugal}
\affiliation{DCTI/ISCTE Instituto Universit\'ario de Lisboa, 1649-026 Lisbon, Portugal}
\author{A.~Martinez de la Ossa}
\affiliation{Institut f\"ur Experimentalphysik, Universit\"at Hamburg, 22761 Hamburg, Germany}
\author{J.~Vieira}\email{jorge.vieira@ist.utl.pt}
\affiliation{GoLP/Instituto de Plasmas e Fus\~ao Nuclear, Instituto Superior T\'ecnico, Universidade de Lisboa, 1049-001 Lisboa, Portugal}

\date{\today}	

\begin{abstract}
Current models predict the hose instability to crucially limit the applicability of plasma-wakefield accelerators.
By developing an analytical model which incorporates the evolution of the hose instability over long propagation distances, this work demonstrates that the inherent drive-beam energy loss, along with an initial beam energy spread detune the betatron oscillations of beam electrons, and thereby mitigate the instability. 
It is also shown that tapered plasma profiles can strongly reduce initial hosing seeds. 
Hence, we demonstrate that the propagation of a drive beam can be stabilized over long propagation distances, paving the way for the acceleration of high-quality electron beams in plasma-wakefield accelerators. We find excellent agreement between our models and particle-in-cell simulations.
\end{abstract}

\maketitle
\emph{Introduction} - 
Plasma-based accelerators can provide accelerating fields in excess of $10$ GV/m \cite{Modena:1995,Blumenfeld:2007} and hence are considered a technology candidate capable of leveraging a dramatic miniaturization of future accelerators and preventing the current scientific progress from faltering in terms of provided beam energy, versatility and availability of accelerator facilities. 
Plasma-wakefield accelerators (PWFA) \cite{Veksler:1956,Chen:1985} employ charged particle beams as drivers of large amplitude plasma waves.
Significant experimental results \cite{Blumenfeld:2007,Litos:2014} were obtained in the \emph{blowout regime}, in which a particle beam with a charge density greater than the ambient plasma density expels all plasma electrons within its vicinity, thereby generating a co-propagating ion-channel with linear electron focusing and extreme accelerating fields \cite{Rosenzweig:1991}.

Identified by D.~Whittum et al.~in the early 1990's \cite{Whittum:1991}, the \emph{hose instability} remains a long standing challenge for PWFA.
Hosing is seeded by initial transverse asymmetries of the beam or plasma spatial or momentum distributions.
According to current models, the beam centroid displacement is amplified exponentially during the beam propagation in the plasma \cite{Whittum:1991, Lampe:1993, Geraci:2000,Deng:2006,Huang:2007}, resulting in an unstable acceleration process or in beam-breakup.
The most recent description for the coupled evolution of the ion-channel centroid $X_c(\xi,t)$ and the beam centroid $X_b(\xi,t)$ in the blowout regime is given by \cite{Huang:2007}
\begin{align}
\frac{\partial^2 X_c}{\partial \xi^2} + \frac{k_p^2 c_\psi(\xi) c_r(\xi)}{2} \left(X_c - X_b\right)&= 0~, \label{eq:huang} \\
\frac{\partial^2 X_b}{\partial t^2} + \omega_\beta^2 \left( X_b - X_c \right) &= 0 ~,\label{eq:mono-ene-ode}
\end{align}
with the time $t$, the co-moving coordinate $\xi=ct-z$, and where $z$ is the longitudinal coordinate and $c$ is the speed of light. The plasma wavenumber is denoted by $k_p=\omega_p/c$, and the betatron frequency by $\omega_\beta=\omega_p/\sqrt{2\gamma}$, with the Lorentz factor $\gamma$, and where $\omega_p=\sqrt{4\pi n_0 e^2/m}$ is the plasma frequency with the ambient plasma density $n_0$, the elementary charge $e$ and the electron rest mass $m$.
The coefficients $c_\psi(\xi)$ and $c_r(\xi)$ account for the relativistic motion of electrons in the blowout sheath and for a $\xi$-dependence of the blowout radius and the beam current \cite{Huang:2007}. According to Eq.~(\ref{eq:huang}), a beam centroid displacement $X_b$ leads to a displacement of the ion-channel centroid $X_c$ along the beam. The displacement $X_c$ then couples back to the temporal evolution of $X_b$ according to Eq.~(\ref{eq:mono-ene-ode}).
The case where $c_\psi=c_r=1$ recovers the seminal hosing model~\cite{Whittum:1991}. This limit, which accounts for a linear response of sheath electrons, is characterised by an exponential growth of $X_b$ and $X_c$ with increasing $\xi$ and $t$ \cite{Lampe:1993, Geraci:2000}.
Owed to the nonlinear response of the electron sheath, the growth rates decrease in the blowout regime, because $c_\psi c_r < 1$~\cite{Huang:2007}.
Despite this reduction of the growth rate, however, the current theoretical descriptions still predict that hosing eventually results in beam breakup during the propagation, and hence poses a strong constraint for the applicability of PWFAs.
\begin{figure}[ht]
	\centering
	\includegraphics[width=1.0\columnwidth]{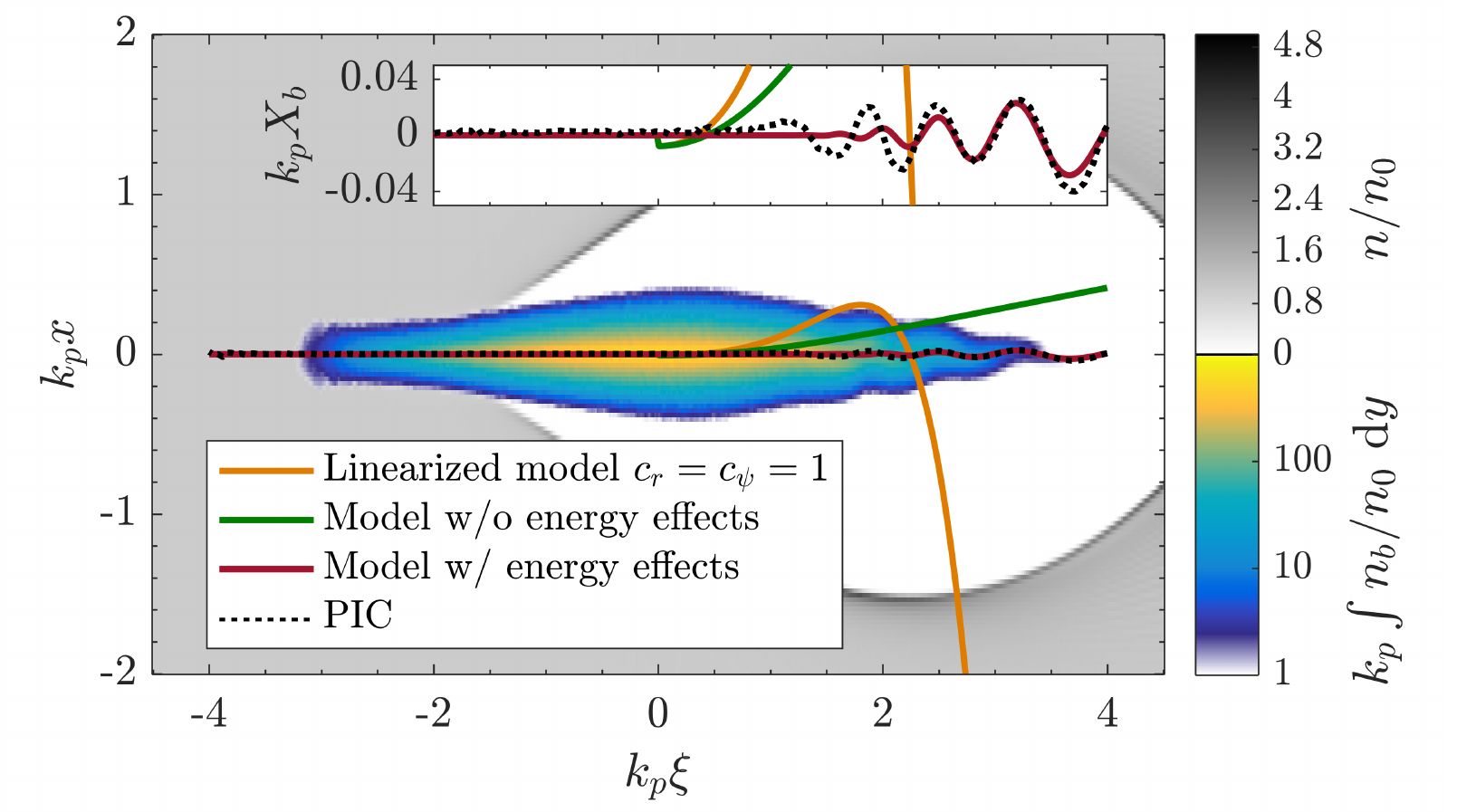}
	\caption{Result from a 3D PIC simulation showing plasma and beam charge densities at time $\omega_{\beta,0} t = 71.6$. The beam has an initial spatial centroid offset, introduced at position $\xi=0$, and is subject to hosing. Beam charge density $n_b$ is projected onto the shown $x$-$\xi$ plane. Lines indicate $X_b(\xi)$, as a result from the models in Ref.~\cite{Whittum:1991, Lampe:1993} (orange solid) and Ref.~\cite{Huang:2007} (green solid), respectively. Depicted is also the result from Eq.~(\ref{eq:huang}) and Eq.~(\ref{eq:beam_centroid_ode}), derived within this work (red solid), and $X_b(\xi)$ retrieved from the PIC simulation (black dashed). Inset: Enlarged depiction of the beam centroids.}
\label{fig:fig1} 
\end{figure}

Although in agreement with particle-in-cell (PIC) simulations for short propagation distances~\cite{Huang:2007}, current models overestimate the hosing growth rates as soon as the drive-beam energy change becomes significant. This is shown in Fig.~\ref{fig:fig1}, which depicts the result of a three-dimensional PIC simulation with OSIRIS~\cite{Fonseca:2002, *Fonseca:2008, *Fonseca:2013}, indicating that hosing can be far less pronounced than what has been reported so far. Yet unnoticed, this intriguing result suggests that the blowout regime can provide a saturation mechanism for the hose instability, which strongly damps the beam centroid oscillations during the propagation, thereby contributing to the stabilization of the beam propagation over long distances.

\begin{figure}[t]
	\centering
	\includegraphics[width=1.0\columnwidth]{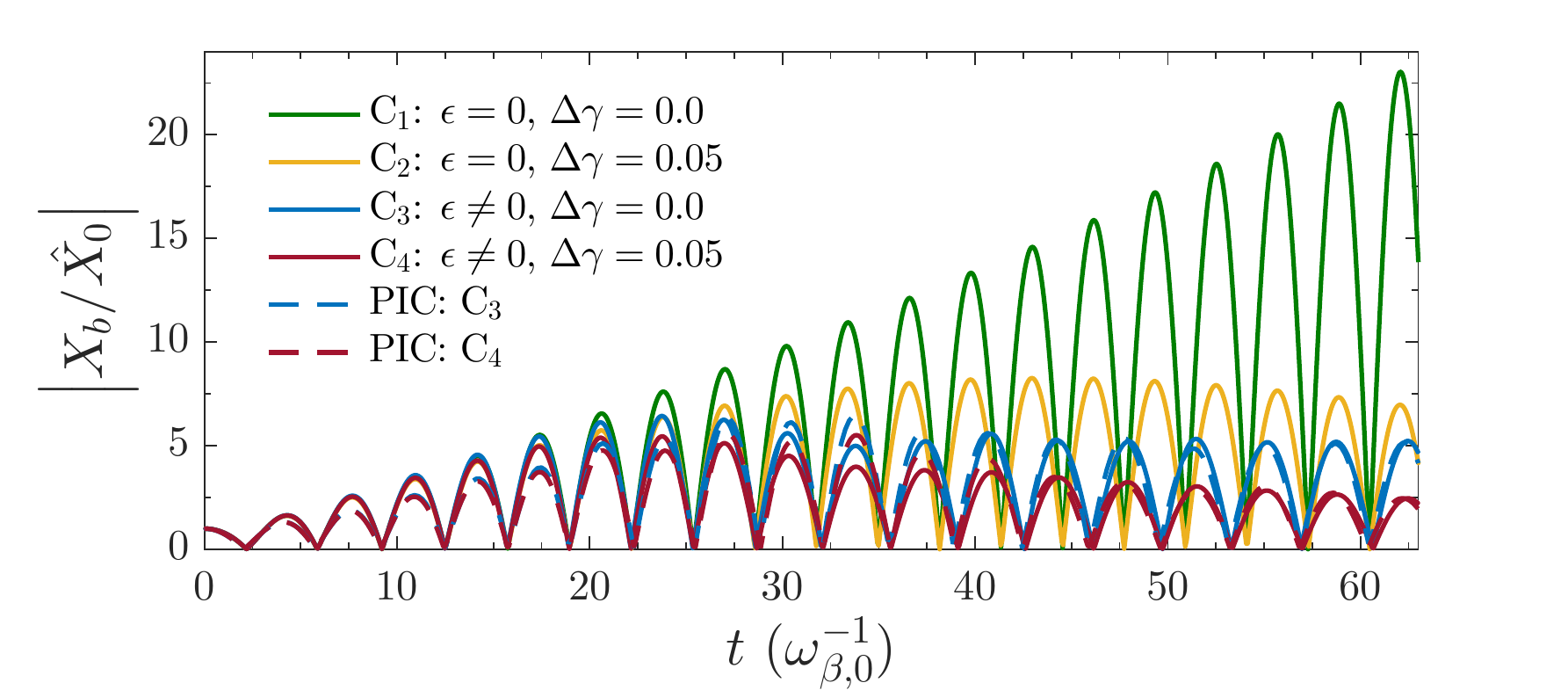}
	\caption{Absolute value of the beam centroid at $k_p\xi=3.0$. Depicted are numerical solutions of Eq.~(\ref{eq:beam_centroid_ode}) for no energy change $\epsilon = 0$ and no energy spread $\Delta \gamma =0.0$ (green solid), for $\epsilon = 0$ and $\Delta \gamma =0.05$ (yellow solid), for $\epsilon \neq 0$ and $\Delta \gamma =0.0$ (blue solid) and for $\epsilon \neq 0$ and $\Delta\gamma = 0.05$ (red solid). These curves are compared to the results of PIC simulations (dashed).}
\label{fig:fig2} 
\end{figure}

In this Letter we show by means of analytical theory and with PIC simulations that hosing can be mitigated in the blowout regime. This has not previously been identified because current analytic models neglect the energy change of the drive-beam particles. Instead, here we find that the energy change, which naturally occurs as the beam excites the plasma wave, and/or an initial beam energy chirp, can detune the betatron oscillations of individual slices along the beam, thereby mitigating their resonant coupling via the plasma. We also show that beam-centroid oscillations can significantly be reduced if the drive-beam features a sub-percent uncorrelated energy spread, which introduces a decoherence of the betatron oscillations of individual beam electrons. Our theoretical model can accurately explain the reduced centroid amplitude of oscillations observed in simulations, as shown in Fig.~\ref{fig:fig1} (see dashed line and solid red line) and in Fig.~\ref{fig:fig2}. We also propose to substantially decrease the initial hosing seed by using tailored vacuum-to-plasma transitions. 
This letter outlines these physical phenomena limiting the detrimental effects of hosing. 
We confirm all our analytical predictions with three-dimensional PIC simulations using OSIRIS. 
The parameters used in these simulations differ from those proposed for a number of high-energy beam facilities. 
Numerical demonstrations of the hosing saturation for parameters corresponding to these facilities are to be published elsewhere
\footnote{To be submitted.}.
Our findings pave the way for stable acceleration of high-quality beams over long distances in PWFAs, and provide theoretical evidence for why hosing to date has not been experimentally detected.

\emph{Derivation of the beam-centroid equation} - 
The starting point is the differential equation for the transverse position $x$ of a single beam electron relative to the axis in a homogeneous ion-channel \cite{Glinec:2008, Vieira:2011b}
\begin{equation}
\frac{d^2x}{dt^2} + \frac{\dot{\gamma}}{\gamma} \frac{dx}{dt} + \omega_\beta^2 \left( x - X_c \right) = 0~, \label{eq:part_dyn}
\end{equation}
where $\dot{\gamma}=d\gamma/dt$. 
The Lorentz factor $\gamma \simeq p_z/mc \gg 1$, with the longitudinal momentum $p_z$, is decoupled from the transverse motion, since $dx/dt \ll c$. Radiation effects are neglected.
The term $\dot{\gamma}/\gamma$ results in a damping or amplification of the amplitude of the single-electron oscillation, depending on whether the electron gains ($\dot{\gamma}>0$) or loses ($\dot{\gamma}<0$) energy, respectively. 
The restoring force is directed towards the channel centroid $X_c$.
The solution for Eq.~(\ref{eq:part_dyn}) is
\begin{align}
x(t) \simeq &~x_0 A(t) \cos\left[ \varphi(t) \right] + \frac{p_{x,0}}{m\gamma_0 \omega_{\beta,0}} A(t) \sin\left[ \varphi(t) \right] \label{eq:sing_ele_sol} \\
&+ \omega_{\beta,0}\int^{t}_0 A(t) A(t') \sin \left[\varphi(t) -\varphi(t')\right]\, X_c(t')~\mathrm{d}t' ~, \nonumber
\end{align}
where $\omega_{\beta,0}=\omega_p/\sqrt{2\gamma_0}$, $A(t)=[\gamma_0/\gamma(t)]^{1/4}$, and where $\gamma_0$ and $p_{x,0}$ are the initial Lorentz factor and transverse momentum, respectively. The phase-advance is defined by $\varphi(t) = \int \omega_\beta \mathrm{d}t$. The relative energy and amplitude variations occur on timescales longer than the betatron period in relevant scenarios. Thus, the terms $|\dot{\gamma}\dot{A}/(\dot{\varphi}^2 \gamma A ) | \ll 1$, $|\ddot{A}/(\dot{\varphi}^2 A) |\ll 1$ and $|\dot{\gamma} / 4 \gamma_0 \omega_{\beta,0} | \ll 1$ were neglected.

In the following, the energy of an electron is given by $\gamma(t) = \overline{\gamma}_0 + \mathcal{E} t+ \delta\gamma$, where $\overline{\gamma}_0=\overline{\gamma}_0(\xi)$ is the initial mean slice energy as a function of the co-moving coordinate, accounting for an initial energy chirp. The differential change of energy along the beam is accounted for by means of the term $\mathcal{E} t$, where $\mathcal{E}=-eE_z/m c$, where $E_z=E_z(\xi)$ is the longitudinal electric field, and where electrons are fixed to their initial position in the co-moving frame.
The uncorrelated energy spread is incorporated through a finite deviation of the electron energy from the mean slice energy $\delta\gamma =\gamma - \overline{\gamma}$.
All overlined quantities refer to slice-averaged quantities.

Electrons with a small relative energy deviation $ | \delta\gamma/\overline{\gamma} | \ll 1$ have a betatron frequency $\omega_\beta$ which deviates from $\overline{\omega_{\beta}}$ according to $\omega_\beta \simeq \overline{\omega_\beta} (1 - \delta\gamma/2\overline{\gamma})$. Hence: 
\begin{equation}
\varphi(t) = \overline{\varphi}(t) \left( 1 - \frac{\delta\gamma}{2\overline{\gamma_0}} \frac{\overline{\omega_\beta}}{\overline{\omega_{\beta,0}}} \right) ~, \label{eq:phase_adv}
\end{equation}
where $\overline{\varphi} = 2(\overline{\omega_{\beta,0}}/\overline{\omega_{\beta}}-1)/\epsilon$, $\overline{\omega_{\beta,0}}=\omega_p/\sqrt{2\overline{\gamma_0}}$ and $\overline{\omega_{\beta}} = \overline{\omega_{\beta,0}}/\sqrt{1+\epsilon \overline{\omega_{\beta,0}} t }$.
Note that $\overline{\omega_{\beta}}$ is time-dependent owing to a finite relative energy change per betatron cycle $\epsilon = \mathcal{E}/\overline{\gamma_0} \, \overline{\omega_{\beta,0}} = -\sqrt{2/\overline{\gamma_0}} ~ E_z/E_0$, with $E_0=\omega_p m c/e$. Eq.~(\ref{eq:phase_adv}) infers that electrons with differing energy within a slice acquire a differing phase advance, which leads to the phase-mixing of the electron betatron oscillations. This phase mixing can damp the hose instability, similarly to the damping of the hosing of fully self-modulated beams through a change of the betatron frequency \cite{Vieira:2014}.

In order to assess the effect of the phase-mixing onto the hose instability, the beam centroid $X_b$ at a given co-moving position $\xi$ is deduced from Eq.~(\ref{eq:sing_ele_sol}) by averaging with respect to an initial phase-space distribution $f_0(x_0,p_{x,0},\gamma_0)$ within each beam slice,
$X_{b}(\xi,t) =\int x\,f_0 \, \mathrm{d}x_0\,\mathrm{d}p_{x,0}\,\mathrm{d} \gamma_0$,
with $\int f_0\,\mathrm{d}x_0\,\mathrm{d}p_{x,0}\,\mathrm{d} \gamma_0 = 1$.
We assume that the initial transverse offset and momentum in a slice are not correlated with energy. Hence, $f_0$ is separable $f_0=f_\perp(x_0,p_{x,0}) ~f_\gamma(\gamma_0)$. 
While the distribution $f_{\perp}(x_0,p_{x,0})$ is arbitrary (apart from assuming $f_{\perp}=0$ outside the channel) with a mean spatial value $\overline{x_0} = X_{b,0}$, the energy distribution considered here complies with a Gaussian distribution
$f_{\gamma}=(\sqrt{2\pi}\sigma_\gamma)^{-1}\exp\left(-\delta\gamma^2/2\sigma_\gamma^2\right)$.
Averaging over the initial transverse phase space distribution and over the Gaussian energy distribution, neglecting the variation of $A$ owed to $\delta \gamma$, yields
\begin{align}
&X_{b}(\xi,t) \simeq \label{eq:beam_sol} \\
&X_{b,0}(\xi) \overline{A}(\xi,t) \exp \left(-\frac{\Delta\gamma^2 \overline{\alpha}(\xi,t)^2}{2} \right) \cos \left[ \overline{\varphi}(\xi,t) \right] \nonumber \\
&+ \int^t_0 \,\overline{A}(\xi,t) \overline{A}(\xi,t') \exp \left[ -\frac{\Delta\gamma^2(\overline{\alpha}(\xi,t)^2 - \overline{\alpha}(\xi,t')^2)}{2} \right] \nonumber \\ 
&\times\sin\left[\overline{\varphi}(\xi,t) -\overline{\varphi}(\xi,t')\right] X_c(\xi,t') \, \overline{\omega_{\beta,0}}(\xi) \, \mathrm{d}t' \nonumber ~,
\end{align}
with the initial relative energy spread $\Delta\gamma=\sigma_\gamma/\overline{\gamma_0}$, the amplitude $\overline{A}= (\overline{\gamma_0}/\overline{\gamma})^{1/4}$ and $\overline{\alpha} = \overline{\varphi}\, \overline{\omega_\beta}/2\overline{\omega_{\beta,0}}$. The initial mean slice transverse momentum is assumed zero for compactness.
Equations~(\ref{eq:huang}) and (\ref{eq:beam_sol}) describe the coupled evolution of $X_c$ and $X_b$ in the blowout regime. They recover known results in the blowout regime at sufficiently early times~\cite{Huang:2007}. When energy effects become relevant, however, they show that hosing can be mitigated.

\emph{Interpretation using a two-particle beam} - 
In order to investigate the physical predictions of Eqs.~(\ref{eq:huang}) and (\ref{eq:beam_sol}) analytically, we use a two-particle (two-slice) model
such that $X_b(\xi,t)=X_{b,1}(\xi,t) \delta(\xi-\xi_1)+X_{b,2}(\xi,t) \delta(\xi-\xi_2)$. The first slice, at $\xi_1$, is unaffected by the hose instability, but drives the channel centroid oscillations according to Eq.~(\ref{eq:huang}). The motion of the slice at $\xi_2$ is driven by those channel oscillations according to Eq.~(\ref{eq:beam_sol}) \cite{Note1}.

We start by determining the timescale for the hosing mitigation by isolating the contributions of finite $\partial_\xi \epsilon \neq 0$, which accounts for the differential energy change along the beam. Analytical results are valid for arbitrary $c_r(\xi)$, $c_\psi(\xi)$, for beams without initial energy spread, and for constant $\overline{A}$. 
Initially, the trailing slice is resonantly driven by the transverse motion of the first slice, enhancing the amplitude of $X_b(\xi_2)$. This corresponds to the initial hosing growth investigated in Ref.~\cite{Huang:2007}. 
However, at time $\overline{\omega_{\beta,0}} t_{\mathrm{d},\epsilon} \simeq \sqrt{3\pi/\Delta\epsilon}$, where $\Delta\epsilon = |\epsilon(\xi_1)-\epsilon(\xi_2)|$, when the phase difference of the two slices is significant, $X_b(\xi_2)$ reaches a maximum \cite{Note1}.
For $t>t_{\mathrm{d},\epsilon}$, the oscillation amplitude of $X_b(\xi_2)$ saturates at a smaller value. 
This fundamentally novel result is in strong contrast with current models, which predict exponentially growing amplitudes until beam breakup.

This finding is significant because the pump depletion time is typically much longer than $t_{\mathrm{d},\epsilon}$. 
We demonstrate this by comparing the pump depletion time, given by $t_{\mathrm{dp}}=1/\overline{\omega_{\beta,0}}\,\hat{\epsilon}$, to $t_{\mathrm{d},\epsilon}$, where $\hat{\epsilon} = \mathrm{max}(-\epsilon)$.
Hence, decoupling of two slices occurs well before pump depletion if $\Delta\epsilon/\hat{\epsilon} > 3\pi \hat{\epsilon}$. Because $|\hat{\epsilon} |\ll 1$ and since $\Delta\epsilon/\hat{\epsilon}$ ranges from zero to unity along any drive beam, the two particle model suggests that slices within the beam in PWFAs are decoupled significantly before depletion.

The parameter $\epsilon$ is related to key experimental PWFA parameters as follows.
The longitudinal field within the beam region can be approximated by $E_z/E_0\simeq \sqrt{I_b/I_A}$ \cite{Lotov:2004, MartinezDeLaOssa:2015}, where $I_b$ is the beam current and $I_A \simeq 17$ kA is the Alfv\'en current. Hence, $\epsilon \simeq -\sqrt{2 I_b/(I_A \overline{\gamma_0})}$, and FACET experimental parameters \cite{Hogan:2010, Adli:2013}, for instance, yield $\hat{\epsilon} \approx 0.007$. 
This result indicates that the growth of the hose instability stops well before energy depletion in typical PWFA scenarios and possibly justifies why hosing was not detected in previous experiments~\cite{Blumenfeld:2007, Litos:2014}.

The two-particle model also indicates that an initial linear energy chirp, $\chi = \gamma_{b}^{-1} k_p^{-1} d \overline{\gamma}/ d\xi $, can mitigate hosing. The centroid oscillations of two spatially resonant beam slices ($\Delta \xi = k_p^{-1}\pi\sqrt{2}$) decouple after $\omega_{\beta,b} t_{\mathrm{d},\chi} \simeq \sqrt{2}/| \chi|$, assuming $c_r c_\psi=1$ and $\epsilon=0$ \cite{Note1}. The damping due to this effect is similar to BNS damping \cite{Balakin:1983}.
Here, $\gamma_{b}$ and $\omega_{\beta,b}$ refer to the initial beam-averaged Lorentz factor and betatron frequency, respectively.

Additionally, according to Eq.~(\ref{eq:beam_sol}), the amplitude of the $X_b$ oscillations are damped exponentially owed to a finite uncorrelated energy spread. To isolate this effect, we consider a beam with no initial chirp in the limit of no slice energy change ($\epsilon \rightarrow 0$). In this conservative scenario, the amplitude of the centroid oscillations reduces by $\exp(-1/2)$ after the decoherence time $\overline{\omega_{\beta,0}}t_{\mathrm{d},\Delta\gamma} \simeq 2/\Delta\gamma$ \cite{Note1}. Therefore, $t_{\mathrm{d},\Delta\gamma} \lesssim t_{\mathrm{dp}}$ if $\Delta \gamma \gtrsim 2\hat{\epsilon}$.
For the typical parameters of FACET, where $\hat{\epsilon} \approx 0.007$, a sub-percent-level energy spread already significantly contributes to the mitigation of hosing. 
It should be noted that if $t_{\mathrm{d},\epsilon}\lesssim t_{\mathrm{d},\Delta\gamma}$, the exponential damping of $X_b$ due to the uncorrelated energy spread becomes substantial since $X_b$ stops growing owing to finite $\partial_{\xi}\epsilon \ne 0$.

\emph{Numerical results \& comparison to PIC simulations} - 
Because fully analytical solutions of our model are complex, we complement the analysis of the two-particle model with PIC simulations using OSIRIS \cite{Fonseca:2002, Fonseca:2008, Fonseca:2013} and with the numerical solution of Eq.~(\ref{eq:huang}) and the differential form of Eq.~(\ref{eq:beam_sol}), given by:
\begin{equation}
\begin{aligned}
\frac{\partial^2 X_b}{\partial t^2} &+ \frac{\overline{\omega_{\beta}}^2}{\overline{\omega_{\beta,0}}}\left(\epsilon + \kappa_1\Delta\gamma^2 \right) \frac{\partial X_b}{\partial t} \\
&+\overline{\omega_{\beta}}^2 (1+\kappa_2\Delta\gamma^2 ) (X_b-X_c) = 0 ~ , \label{eq:beam_centroid_ode}
\end{aligned}
\end{equation}
with $\kappa_1 = (\overline{\omega_{\beta}}/\overline{\omega_{\beta,0}}-(\overline{\omega_{\beta}}/\overline{\omega_{\beta,0}})^2)/\epsilon$, and $\kappa_2 = (\overline{\omega_{\beta}}/\overline{\omega_{\beta,0}})^4/2-(\overline{\omega_{\beta}}/\overline{\omega_{\beta,0}})^3/4$. This equation, which neglects terms $\mathcal{O}(\Delta\gamma^4)$ and $\mathcal{O}(\epsilon^2)$, applies for any beam and blowout-regime wakefield.

We consider a Gaussian electron beam with $\overline{\gamma_0}=1956.95$, a peak current of $\hat{I}_b = I_A/4$, transverse dimensions of $k_p\sigma_x=k_p\sigma_y = 0.1$, and longitudinal dimension of $k_p\sigma_z=1.0$, traversing a plasma target with a flat-top density $n_0$ and driving a plasma wave in the blowout regime (cf.~Fig.~\ref{fig:fig1}).
The initial centroid along the beam is given by $k_p X_{b,0} (\xi)=0.01 \times \Theta(k_p \xi)$, where $\Theta(x)$ is the Heaviside-step function. The initial centroid offset is introduced from the peak current location at $\xi=0$. The beam has no initial energy chirp.
Following Ref.~\cite{Huang:2007}, $c_r(\xi)=4 I_b(\xi)/I_A(k_p R(\xi))^2$ and $c_\psi(\xi)=1/(1+\psi(\xi))$ in Eq.~(\ref{eq:huang}), as well as $E_z(\xi)$ for Eq.~(\ref{eq:beam_centroid_ode}), are computed numerically according to the model for the blowout regime in Refs.~\cite{Lu:2006, *Lu:2006b,Yi:2013}.
Here, $R$ is the blowout radius and $\psi= (\phi - A_z) e/mc^2$ the normalized wakefield potential in the sheath, with the electrostatic potential $\phi$ and longitudinal vector potential $A_z$. 

Numerical solutions of Eqs.~(\ref{eq:huang}) and (\ref{eq:beam_centroid_ode}) are depicted in Fig.~\ref{fig:fig2} for the cases $\mathrm{C}_1$: $\epsilon=0$, $\Delta\gamma = 0.0$; $\mathrm{C}_2$: $\epsilon=0$, $\Delta\gamma = 0.05$; $\mathrm{C}_3$: $\epsilon\neq0$, $\Delta\gamma = 0.0$; and $\mathrm{C}_4$: $\epsilon\neq0$, $\Delta\gamma = 0.05$, together with results from PIC simulations for the two latter cases ($\mathrm{C}_4$ also corresponds to the result shown in Fig.~\ref{fig:fig1}).
Case $\mathrm{C}_1$, which resembles the model in Ref.~\cite{Huang:2007}, features the expected exponential growth rate, as illustrated in Fig.~\ref{fig:fig2}. 
For $\mathrm{C}_3$, the detuning of the slices betatron oscillations leads to a saturation of the hose instability. According to the two-particle model, the maximum amplitude for $\mathrm{C}_3$ is expected near $\overline{\omega_{\beta,0}} t_{\mathrm{d},\epsilon} \approx 22.7$ ($\Delta \epsilon$ between $k_p\xi=0$ and the depicted slice at $k_p\xi=3.0$), which is in good agreement with the numerical result and the result obtained from the PIC simulation.
Moreover, in $\mathrm{C}_2$ and $\mathrm{C}_4$, the centroid oscillations are damped because of the energy-spread induced betatron decoherence within the slices.
In $\mathrm{C}_2$ and $\mathrm{C}_4$, the energy spread is $\Delta\gamma = 0.05$, thus yielding  $\overline{\omega_{\beta,0}}t_{\mathrm{d},\Delta\gamma} = 40$. 
The corresponding exponential damping of $X_b$ for $t \gtrsim t_{\mathrm{d},\Delta\gamma}$ is in good agreement with the observations in Fig.~\ref{fig:fig2} for both, the numerical solution of Eqs.~(\ref{eq:huang}) and (\ref{eq:beam_centroid_ode}) and the particle-in-cell simulation.

Effective damping of the hose instability can occur as long as the hosing seed is sufficiently small not to lead to beam breakup before the mitigation takes place.
Reducing the initial hose seed is therefore still crucial to fully stabilize the driver propagation.
For this purpose we propose a novel concept which employs plasma density tapers to mitigate initial beam centroid offsets that seed hosing. 

\emph{Mitigation of hosing with plasma-density tapers} - We consider a taper of the plasma density from the vacuum-to-plasma interface at position $z_v$ to the flat-top plasma profile from position $z_0$.
The beam centroid during the propagation in the tailored vacuum-to-plasma transition is described by
\begin{equation}
\frac{d^2 X_b}{dz^2} + k_\beta(z)^2 X_b = 0 ~,\label{eq:taper_beam_dyn}
\end{equation}
with $k_\beta = k_{\beta,0} \sqrt{n/n_0}$, when neglecting the channel centroid displacement, the beam-energy change and effects from energy spread. 
This equation corresponds to the non-conservative system of an harmonic oscillator with time-dependent frequency.
The beam centroid is therefore damped during the propagation through the taper.

To confirm the hosing seed mitigation scheme, we re-ran PIC simulations of $\mathrm{C}_3$ with a tapered plasma density profile.
The considered propagation-distance dependent betatron wavenumber, $k_\beta = \omega_\beta/c$, is given by $k_\beta(z) = k_{\beta,0}(1-(z-z0)/\lambda_{\mathrm{opt}})^{-2}$ for $z_v <z \leq z_0$, $k_\beta(z) = k_{\beta,0}$ for $z > z_0$ and $k_\beta(z) = 0$ otherwise (this functional dependence was used for the beam betatron function matching in Refs.~\cite{Floettmann:2014,Xu:2016}). Here $\lambda_{\mathrm{opt}} \simeq L/\sqrt{k_{\beta,0} L}$ is an optimized characteristic scale length of the taper. 
Such density profiles can be experimentally realized in appropriate gas capillaries \cite{Schaper:2014}.

\begin{figure}[h]
	\centering
	\includegraphics[width=1.0\columnwidth]{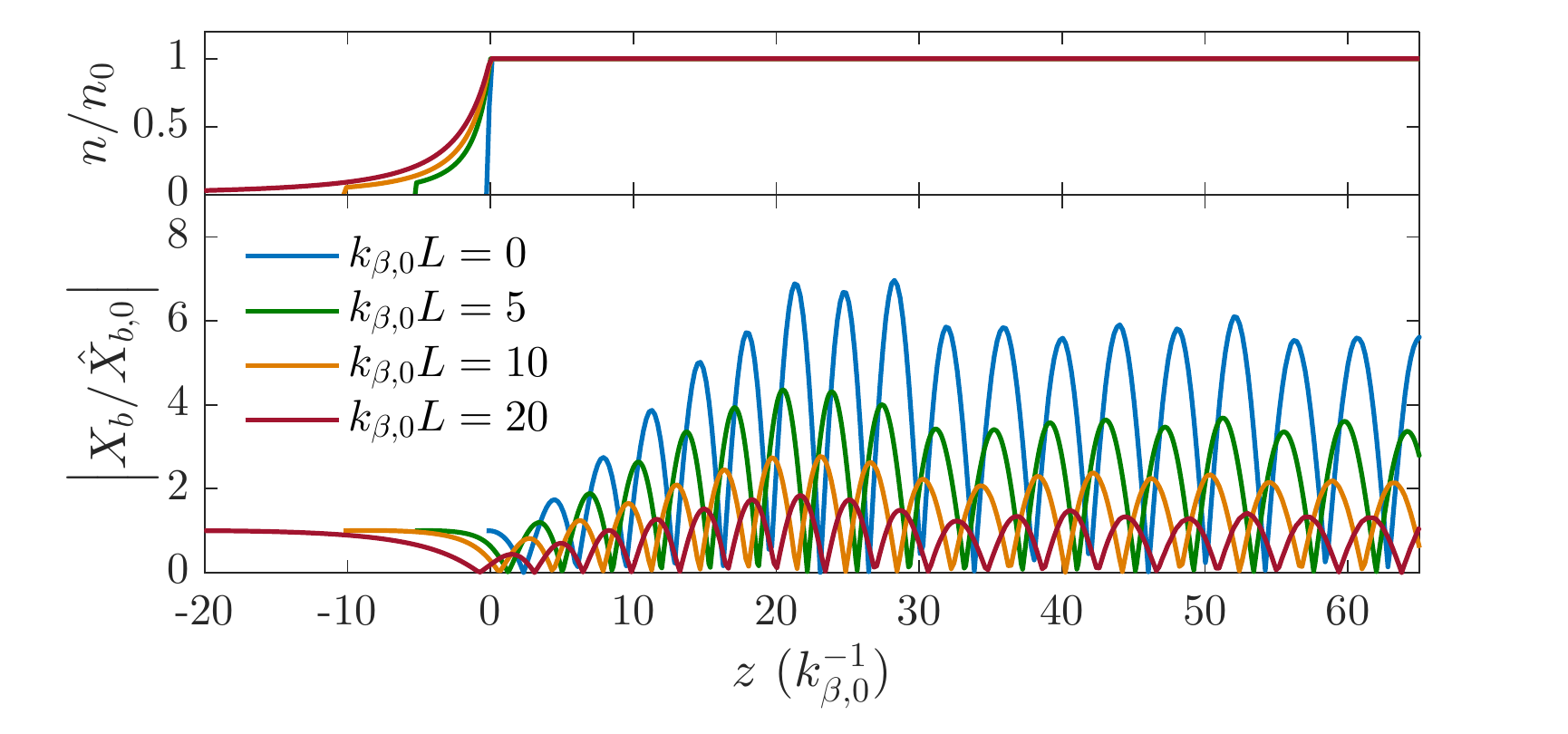}%
	\caption{Hose mitigation by means of plasma density tapers at the tail of a beam $k_p\xi = 4.0$. Shown are density profiles for different taper lengths (top) and respective beam centroid amplitudes from PIC simulations (bottom) for $k_{\beta,0} L=0$ (blue), $k_{\beta,0} L=5$ (green), $k_{\beta,0} L=10$ (orange) and $k_{\beta,0} L=20$ (red).}
\label{fig:fig3}
\end{figure}
The evolution of $X_b$ for various taper lengths are depicted in Fig.~\ref{fig:fig3}, illustrating the substantial reduction of the hose instability when $k_{\beta,0} L \gtrsim 1$, compared to the case with no taper.

\emph{Summary and conclusion} - 
This work demonstrates that the self-consistent beam energy evolution in the blowout regime can mitigate the hose instability in PWFAs.
We show that the drive-beam energy chirp, either introduced initially or developed during propagation, results in the mitigation of the hose instability before pump depletion, regardless of the initial beam energy \cite{Note1}.
We also find that an initially sub-percent uncorrelated energy spread will further reduce the centroid oscillations.
Furthermore, it is shown that tapering the plasma profile can efficiently reduce the initial hose seed.

\begin{acknowledgements}
\emph{Acknowledgments} - We acknowledge the grant of computing time by the J\"ulich Supercomputing Centre on JUQUEEN under Project No.~HHH23 and the use of the DESY IT high-performance computing facilities.
This work was supported by LaserLab Europe IV - grant agreement 654148 (H2020-INFRAIA-2014-2015) and EuPRAXIA - grant agreement 653782 (H2020-INFRADEV-1-2014-1).
The work leading to this publication was supported by the German Academic Exchange Service (DAAD) with funds from the German Federal Ministry of Education and Research (BMBF) and the People Programme (Marie Curie Actions) of the European Union's Seventh Framework Programme (FP7/2007-2013) under REA grant agreement no.~605728 (P.R.I.M.E.~- Postdoctoral Researchers International Mobility Experience). 
\end{acknowledgements}

\end{document}